\documentclass[12pt]{article}
\usepackage{graphicx}
\usepackage{bm}
\usepackage{amsmath,amssymb,amsfonts}
\newcommand{\Slash}[1]{\ooalign{\hfil/\hfil\crcr$#1$}}

\title{Photon induced $\Lambda(1520)$ production and \\the role of the $K^*$ exchange}
\author{Hiroshi Toki$^{a,b}$, Carmen Garc\'{\i}a-Recio$^a$ and 
Juan Nieves$^a$\\
a) Departamento de F\'\i sica  At\'omica, Molecular y Nuclear,
\\Universidad de Granada, E-18071, Spain\\
b) Research Center for Nuclear Physics (RCNP), \\ Osaka University, 
Ibaraki, Osaka 567-0047, Japan
}

\begin{document}
\maketitle

\date

\abstract{ We study the photon induced $\Lambda(1520)$ production in
  the effective Lagrangian method near threshold, $ E_\gamma^{\rm LAB}
  \le 2$ GeV, and in the quark-gluon string model at higher energies 3
  GeV $\leq E_\gamma^{\rm LAB} \leq 5$ GeV.  In particular, we study
  the role of the $K^*$ exchange for the production of $\Lambda(1520)$
  within the SU(6) Weinberg-Tomozowa chiral unitary model proposed in
  Ref.~\cite{garcia06-su6}.  The coupling of  the $\Lambda(1520)$
  resonance to the $N \bar K^*$ pair, which is dynamically
  generated, turns out to be relatively small and, thus, the $K$
  exchange mechanism dominates the reaction.  In the higher energy
  region, where experimental data are available, the quark-gluon
  string mechanism with the $K$ Regge trajectory reproduces both the energy
  and the angular distribution dependences of the $\Lambda(1520)$
  photo-production reaction.}

\section{Introduction}

The recent announcement of the finding of the exotic hyperon, so
called pentaquark, opened a new field for nuclear and particle
physicists to study composite objects with more than three quarks
\cite{nakano03,stepanyan03,airapetian04}.  A correct understanding of
the experimental findings requires to possess  a suitable description
of hadron and photon induced reactions in a region where both, the
baryon-meson and the quark-gluon degrees of freedom, should be
considered, due to the energy and momentum transfers involved. On the
other hand, in the recent years, our knowledge of the structure and
dynamics of s-wave and d-wave odd parity baryon resonances has been
increased thanks to the use of chiral unitarity schemes, which
successfully generate those resonances lying near the composite hadron
threshold energies~\cite{kaiser95,oset98,oller01,LK02,nieves01,garcia04}.  All
these developments make this new research field exciting and
worthwhile for further study.

The claimed pentaquark ($\Theta^+$) has strangeness $S=+1$, zero
isospin, a mass of around 1530 MeV and its spin--parity has not been
identified yet.  Definitely, all these details have to be determined
and even its existence has to be investigated by further detailed
experiments~\cite{nakano07}.  Due to this situation, a lot of
theoretical attention is being paid to the study of the isoscalar
$\Lambda(1520)$ resonance (hereafter called $\Lambda^*$), with similar
mass to that of the pentaquark, but with opposite strangeness
($S=-1$).  In particular, it clearly shows up in the $K^- p$ invariant
mass distribution of the two step process
\begin{equation}
\gamma p \to K^+ \Lambda^* \to K^+ K^- p
\end{equation}
Note that the above reaction has a clear resemblance
to the $\gamma n \to K^- \Theta^+(1530) \to K^- K^+n$ one, where the
pentaquark should appear in the $K^+ n$ mass spectrum\footnote{The
LEPS collaboration~\cite{nakano03} uses a deuterium target since
neutrons are unstable particles.}. The
$\Lambda^*$ spin-parity is $J^\pi=3/2^-$, it lies slightly below the
threshold energy of the $\pi \Sigma^*(1380)$ channel and it decays
into a d-wave anti-kaon nucleon pair.  The recent extended SU(3)
chiral unitarity model of Ref.~\cite{roca06}, which involves baryon
decuplet and meson octet degrees of freedom, seems to reasonably
describe the dynamics of this resonance.

There exist several effective hadron Lagrangian
studies~\cite{nam05,titov05} of the $\gamma p \to \Lambda^* K^+$
reaction for laboratory photon  energies ranging from threshold, 
$\frac{(m_{K}+M_{\Lambda^*})^2-M^2_N}{2M_N} \approx $ 1.7~GeV, up to
about 5 GeV, where experimental measurements are available. These
theoretical studies are hampered by the lack of knowledge on the $\bar
K^*N \Lambda^*$ coupling strength.  This fact, in conjunction with the
use of largely different form-factors to account for the compositeness
of the hadrons, has led to contradicting predictions of the dominant
reaction mechanism in the $\gamma p \to \Lambda^* K^+$ reaction. Thus,
Nam et al.~\cite{nam05} claimed that the kaon exchange provides the
leading contribution in the whole energy region, while within the
quark-gluon string model of Titov et al.~\cite{titov05}, the vector
kaon ($K^*$) exchange turns out to be the dominant mechanism.  This
quark--gluon string picture was introduced by
Kaidalov~\cite{kaidalov82}, Donnachie and Landshoff~\cite{donnachie87}
more than twenty years ago, and it has been used for photon and hadron
induced reactions with energies above a few GeV~\cite{grishina06}.

On the other hand, recently, a consistent SU(6) extension of the
Weinberg-Tomozawa (WT) SU(3) chiral Lagrangian has been derived by
Garc\'{\i}a-Recio et al.~\cite{garcia06-su6}. In this manner, the
lowest-lying meson vector nonet and baryon $3/2^+$ decuplet hadrons
are considered in addition to the members of the pion and nucleon
octets originally included in the WT SU(3) interaction term. The
potentials deduced from this SU(6) Lagrangian are used to solve the
coupled channel Bethe Salpeter Equation (BSE), within the so-called on
shell Renormalization Scheme (RS), leading to unitarized s-wave
meson--baryon scattering amplitudes. In what follows we will refer to
this model as $\chi$SU(6)-BSE. The $\chi$SU(6)-BSE model reproduces
the essential features of previous
studies~\cite{kaiser95,oset98,oller01,LK02,nieves01,garcia04} (for
instance, properties of the lowest lying $J^\pi=1/2^-$ and $3/2^-$
resonances, see for instance \cite{garcia07, garcia07b}) and, in
addition, it sheds some light on the role played by the vector mesons
in these processes.

The model assumes that the quark interactions are spin and SU(3)
flavor independent~\cite{gursey64}. This corresponds to treating the
six states of a light quark ($u$, $d$ or $s$ with spin up, $\uparrow$,
or down, $\downarrow$) as equivalent. To speak meaningfully of SU(6)
transformations affecting spin but not orbital angular momentum ($L$)
as invariances, it must be assumed that the orbital angular momentum
and the quark spin are to a good approximation, separately
conserved. This, in turn requires the spin--orbit, tensor and
spin--spin interactions between quarks to be small, which seems to be
the case in the baryon spectrum.  Indeed, SU(6) symmetry in the baryon
sector gets some support from the large $N_c$ limit of QCD, and it
provides several predictions (relatively closeness of baryon octet and
decuplet masses, the axial current coefficient ratio $F/D=2/3$, the
magnetic moment ratio $\mu_p/\mu_n=-3/2$) which are remarkably well
satisfied in nature.

In the meson-baryon language, the fundamental ingredients are the
mesons belonging to the $J_\pi=0^-$ pseudoscalar octet and the
$J_\pi=1^-$ vector nonet SU(3) representations, and the baryons of the
$J^\pi=1/2^+$ nucleon octet and of the $J^\pi=3/2^+$ $\Delta$
decuplet.  In this model, the physical masses and the physical decay
constants are the only SU(6) breaking terms.  After having fixed the
RS, the $\chi$SU(6)-BSE model predicts, up to an overall phase, the
coupling of the $\Lambda^*$ resonance to the different meson-baryon
channels entering in the solution of the SU(6) BSE, and in particular
that to the $\bar K^*N$ pair.  This information will help to fix the
size of $t-$channel $K^*$ exchange contribution to the $\Lambda^*$
photo-production.  In this respect, Hyodo and his
collaborators~\cite{hyodo06} studied the $\bar K^*N \Lambda^*$
coupling using an effective Lagrangian to couple the $K^*$ degrees of
freedom to an extended SU(3) chiral unitarity model, which includes
baryons from the decuplet. They found a rather small value, of the
order of 1, for this coupling constant, while the phenomenological
analysis of the photon induced reaction tends to use larger values.
It is therefore important to find out this coupling in the new
$\chi$SU(6)-BSE scheme, since it provides a consistent unitary chiral
approach which involves also vector mesons.

The purpose of this paper is twofold.  First to calculate the $\bar
K^* N$ pair coupling to the $\Lambda^*$ resonance and second to study
the $\gamma p \to \Lambda^* K^+$ reaction with the new information on
the role play by the $K^*$.  For the latter one, we will use a hybrid model
which combines both the hadron effective Lagrangian approach, for
energies close to threshold, and the quark-gluon string reaction
mechanism  at higher energies.

The paper is arranged as follows.  In Sect.~\ref{sec:rm}, we shall
discuss the effective Lagrangian model (subsection \ref{sec:effect}),
the quark-gluon string approach (subsection \ref{sec:sqg}) and the
hybrid hadron and Reggeon exchange model (subsection \ref{sec:hybrid})
for the photon induced reaction $\gamma p \to \Lambda^* K^+$.  In
Sect.~\ref{sec:su6}, we describe the SU(6) model. Our results are
presented in Section~\ref{sec:res}. In the subsection \ref{sec:gknl},
we extract the $\bar K^*$-nucleon coupling to the $\Lambda^*$
resonance. In the next subsection (\ref{sec:res-sigma}), we discuss
our results for the $\Lambda^*$ photo-production from both the
effective baryon-meson method and the quark-gluon string approach, and
also from the hybrid model.  Finally, in Sect.~\ref{sec:concl}, we
summarize the present study.

\section{Reaction mechanisms }
\label{sec:rm}
In this section we study the reaction 
\begin{equation}
\gamma p \to \Lambda^* K^+
\end{equation}
whose total and angular differential cross sections were measured with
a tagged photon beam (2.8 $ <  E_\gamma < 4.8$ GeV) using the LAMP2
apparatus at the 5 GeV electron synchrotron NINA at Daresbury~\cite{barber80}.

As discussed in the introduction, we will examine two different
approaches based on hadron and quark-gluon degrees of freedom,
respectively, and a hybrid model based on both of them.

\subsection{Effective hadron Lagrangian  method} 
\label{sec:effect}
We begin with the effective Lagrangian approach, and we first list all
the necessary Lagrangian densities (we are just considering charged
nucleon and kaon fields, which are the only ones appearing in the tree
level Feynmman diagrams; in what follows and for simplicity we will
omit any explicit reference to  charges when referring to coupling
constants and masses),
\begin{equation}
\mathcal{L}_{\gamma KK} = -ie (K^- \partial^{\mu} K^+ -
K^+\partial^{\mu}  K^- ) A_{\mu} \label{eq:eqin}  
\end{equation}
\begin{equation}
\mathcal{L}_{Kp\Lambda^*} =
\frac{g_{KN\Lambda^*}}{m_{K}}\bar{\Lambda}^{*\mu}
(\partial_{\mu} K^-){\gamma}_{5}p\,+{\rm h.c.} \label{eq:knl}
\end{equation}
\begin{equation}
\mathcal{L}_{\gamma pp}=-e\bar{p}\left(\Slash{A}-\frac{\kappa_{p}}{2M_{N}}
\sigma_{\mu\nu}(\partial^{\nu}A^{\mu})\right) p + {\rm h.c.}
\end{equation}
\begin{equation}
\mathcal{L}_{\gamma Kp\Lambda^*} = 
-ie\frac{g_{KN\Lambda^*}}{m_{K}}\bar{\Lambda}^{*\mu}
A_{\mu} K^-{\gamma}_{5}p\,+{\rm h.c.} \label{eq:eqfin}
\end{equation}
where $e=\sqrt{4\pi\alpha} > 0$, $\kappa_p$ and $A_\mu$ are the proton
charge, magnetic moment and photon field, respectively, and we use an
obvious notation for the hadron masses, coupling constants and
fields\footnote{We use a Rarita-Schwinger field to describe the
$\Lambda^*$ resonance, $K^+ = (K^-)^\dagger$ annihilates a $K^+$ or
creates a $K^-$ meson and $p$ destroys a proton and
$\alpha=1/137.036$ is the fine-structure constant}.

\begin{figure}[t]
\begin{center}
\includegraphics[scale=0.5]{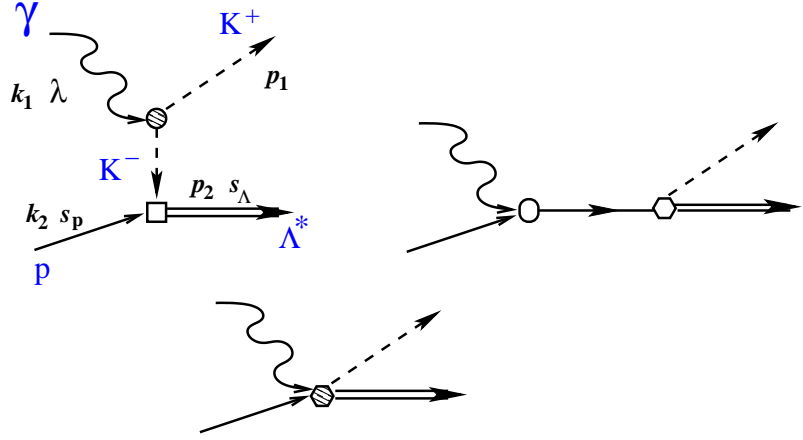}
  \caption{\small $\gamma p \to \Lambda^* K^+$ tree level hadron
  mechanisms constructed out of the Lagrangian densities given in
  Eqs.~(\ref{eq:eqin})--(\ref{eq:eqfin}). In the first diagram we also
  show our definition of the kinematical ($k_1, k_2, p_1, p_2$) 
  and polarization variables ($\lambda, s_\Lambda, s_p$). In
  addition we use $q=k_1-p_1$.}
  \label{fig:dia}
\end {center}
\end{figure}

With the above
Lagrangians one can construct three tree level amplitudes: i)
$t-$channel kaon exchange term, ii) $s-$channel nucleon pole term and
iii) contact term, which are depicted in Fig.~\ref{fig:dia}. All
contributions together provide a gauge invariant amplitude. 
 We do not consider in this work the $u$-channel hyperon pole term
which is by itself gauge invariant and related with the magnetic
coupling of the photon with the $\Lambda^*$, whose information is
scarce~\cite{nam05,titov05}. 

Next we consider the $t-$channel $K^*$ exchange
contribution\footnote{This contribution is readily obtained from the
  first diagram of Fig.~\ref{fig:dia} by replacing the exchanged kaon
  by a vector $K^*$ meson.}, we will need the Lagrangian densities,
\begin{equation}
\mathcal{L}_{\gamma K K^*}=e\frac{g_{\gamma K K^*}}{m_K{^*}}\epsilon^{\alpha \beta \mu \nu}
(\partial_\alpha A_\beta) (\partial_\mu K_\nu^{*+})K^- +{\rm h.c.}
\label{eq:k*rad}
\end{equation}
\begin{eqnarray}
\mathcal{L}_{K^* p \Lambda^*} &=& {\rm i}\, g_{K^* N \Lambda^*}
 \bar \Lambda^*_{\mu} 
 (K^{*-})^\mu p  + {\rm h.c.} \label{eq:swaveK*N}
\end{eqnarray}
with $\epsilon_{0123}=+1$. The above $K^* p \Lambda^*$ vertex is
  predominantly $s-$wave, and its  coupling constant, $g_{K^* N
  \Lambda^*}$ is not known. Below, we will use the $\chi$SU(6)-BSE of
  Ref.~\cite{garcia06-su6} to fix it. Thanks to the transverse nature
  of the $\mathcal{L}_{\gamma K K^*}$ vertex, the $t-$channel $K^*$
  exchange term is gauge invariant by itself. To compare with the
  works of Nam et al.~\cite{nam05} and Titov et al.~\cite{titov05}, we
  have also considered a vector coupling of the $\bar K^*N$ pair to
  the $\Lambda^*$ resonance, which contains a $d-$wave contribution, 
\begin{equation}
\mathcal{L}_{K^* p \Lambda^*}' = \frac{g'_{K^* N
    \Lambda^*}}{m_{K^*}}\bar\Lambda^{*\mu} \gamma^\nu
(\partial_\mu K^{*-}_\nu-\partial _\nu  K^{*-}_\mu)p + {\rm h.c.} 
\label{eq:dwaveK*N}
\end{equation}
where the coupling constant $g'_{K^* N \Lambda^*}$ is not known
either. We denote it with an extra prime ($g'_{K^* N \Lambda^*}$) to
distinguish it from the s-wave coupling used in
Eq.~(\ref{eq:swaveK*N}).  All the other coupling constants can be
fixed from the study of the $K^*$ and $\Lambda^*$ decay widths.

The contribution of the different terms of Fig.~\ref{fig:dia},
including also the $t-$channel $K^*$ exchange, to the $T$-matrix reads
(kinematical and spin variables are explicited in  the first diagram
of Fig.~\ref{fig:dia}) 
\begin{equation}
-iT_i=\bar u_\mu(p_2,s_\Lambda) A_i^{\mu \nu} u(k_2,s_p)
\epsilon_\nu(k_1,\lambda)
\end{equation}
where $u_\mu$ and $u$ are dimensionless Rarita-Schwinger  and  Dirac
spinors, respectively, while $\epsilon_\nu(k_1,\lambda)$ is the photon
polarization vector. 
The reduced  $A_i^{\mu\nu}$ amplitudes related to $t-$channel $K$
exchange are given by
\begin{eqnarray}
A_t^{\mu\nu}&=&-e\frac{g_{KN\Lambda^*}}{m_{K}} \frac{1}{q^2-m_K^2} 
q^\mu (q^\nu - p_1^\nu)\gamma_5\, f_{\rm c} \label{eq:at} 
     \\
A_s^{\mu\nu}&=&-e\frac{g_{KN\Lambda^*}}{m_{K}} \frac{1}{s-M_N^2}\, p_1^\mu 
    \gamma_5 \left(\Slash k_1\, f_{\rm s} +(\Slash k_2+M_N)\, f_{\rm c} \right)
     \gamma^\nu \nonumber \\
&&-e\frac{g_{KN\Lambda^*}}{m_{K}} \frac{1}{s-M_N^2}\, p_1^\mu 
    \gamma_5 (\Slash k_1 +\Slash k_2+M_N) {\rm
      i}\,\frac{\kappa_p}{2M_N}\sigma_{\nu\rho}k_1^\rho\, f_{\rm s}     \\
A_c^{\mu\nu}&=&e\frac{g_{KN\Lambda^*}}{m_{K}}  g^{\mu\nu} \gamma_5\,
f_{\rm c} \label{eq:ac} 
\end{eqnarray}
Here, the momentum transfer carried by the intermediate $\bar K$ is
$q=k_1-p_1$ and the Mandelstam variable $s$ is defined as usual
$s=(k_1+k_2)^2$.  The subindices $t,s$ and $c$ stand for the
$t-$channel kaon exchange, the $s-$channel nucleon pole and the
contact terms, respectively, and are depicted in Fig.~\ref{fig:dia}. The
form-factors $f_{\rm c}$ and $f_{\rm s}$ will be discussed below. We define
\begin{equation}
T_K=T_t+T_s+T_c,
\end{equation}
and in what follows we will refer to it as  the $K$ mechanism
contribution to the $T-$matrix. 
Similarly, for the $K^*$ contribution we get
\begin{equation}
A_v^{\mu\nu}= -e\frac{g_{\gamma K K^*}}{m_K{^*}}
\frac{g_{K^* N \Lambda^*}}{q^2-m_{K^*}^2} 
 \epsilon^{\alpha \nu \beta \mu } k_{1\alpha} q_\beta\,f_{\rm v}
\label{eq:kss} 
\end{equation}
For comparison with previous studies, we also write this contribution
for the case of the vector coupling of Eq.~(\ref{eq:dwaveK*N})
\begin{equation}
A_{v'}^{\mu\nu} = - e\frac{g_{\gamma K K^*}}{m_K{^*}} \frac{g'_{K^* N
\Lambda^*}}{q^2-m_{K^*}^2} k_{1\alpha} q_\beta \left (
\epsilon^{\alpha \nu \beta \mu } \frac{\Slash{q}}{m_{K^*}}
-\frac{q^\mu}{m_{K^*}} \epsilon^{\alpha \nu \beta\sigma} \gamma_\sigma
\right ) \,f_{\rm v}
 \label{eq:ksp}
\end{equation}
where here again the form-factor $f_{\rm v}$ will be discussed below. We
remind here that $\Slash{q}=M_{\Lambda^*}-M_N$, and $-q^\mu
u_\mu(p_2,s_\Lambda) = k_2^\mu u_\mu(p_2,s_\Lambda)$ for on-shell
baryons.  The first term in Eq.~(\ref{eq:ksp}) is of the type of the
contribution in Eq.~(\ref{eq:kss}), and it is plausible to expect
similar values for $g'_{K^* N \Lambda^*}$ and $g_{K^* N \Lambda^*}$,
since $(M_{\Lambda^*}-M_N)/M_{K^*} \approx 2/3$. The second term in
Eq.~(\ref{eq:ksp}) is generated by a $K^{*-}p$ d-wave coupling to the
$\Lambda^*$ resonance.  In general the interaction of
Eq.~(\ref{eq:dwaveK*N}) contains two independent components. In terms
of multipoles, they are $E1$ and $M2$. In the $E1$ amplitude, the
orbital angular momentum of the decaying channel $K^{*-}p$ is
s-wave, while in $M2$, it is d-wave. The $E1$ ($M2$) amplitude is
dominated by the first (second) term in Eq.~(\ref{eq:ksp}). We expect
that the s-wave coupling will dominate near $\Lambda^*$ threshold,
where all involved three momenta at the hadron vertex are small.  We
will apply the effective Lagrangian method only near the threshold
energy, $\sqrt{s} \approx 2$ GeV, and thus we do not expect sizable
effects arising from the second term in Eq.~(\ref{eq:ksp}). In the
recent work of Hyodo et al.,~\cite{hyodo06}, it is also argumented the
dominance of the s-wave $K^*N\Lambda^*$ component of vertex at low and
intermediate energies.

The $K^*$ mechanism contribution to the $T-$matrix is given 
by one of the above expressions, $T_{K^*}=T_v$ or $T_{K^*}=T_{v'}$.

Up to this point, the $T$-matrix is gauge invariant.  However, we
ought to introduce the compositeness of the hadrons.  This is usually
achieved by including form-factors in the amplitudes in such manner
that gauge invariance is preserved\footnote{For the sake of brevity
and to avoid repeating similar equations, in
Eqs.~(\ref{eq:at})-(\ref{eq:ac}) and
Eqs.~(\ref{eq:kss})-(\ref{eq:ksp}),\\ we have already included 
form-factors. Details are given in what follows.}.  There is no unique theoretical way to introduce
the form-factors, this was discussed at length by Ohta~\cite{ohta89}
and by Haberzettl et al.~\cite{haberzettl98}. We adopt here the scheme
used in the previous works \cite{nam05,titov05}, where the
prescription of Ref.~\cite{haberzettl98} was used.  We take the
following parameterization for the form-factors
\begin{eqnarray}
f_{\rm i} &=&\frac{\Lambda^4}{\Lambda^4+(q_{\rm i}^2-M_{\rm i}^2)^2}, \quad {\rm i}={\rm s,t,v} \\
f_{\rm c} &=& f_{\rm s}+f_{\rm t}-f_{\rm s}f_{\rm t}, \quad {\rm and}~
\left\{\begin{array}{l}q_{\rm s}^2=s,\, q_{\rm t}^2=q_{\rm v}^2= q^2   \cr 
                              M_{\rm s} = M_N, \,M_{\rm t} = m_K,\,
			      M_{\rm v}= m_{K^*}
\end{array}\right. \label{eq:cut-had} 
\end{eqnarray}
 In the expressions of the different contributions to $T_K$ and
 $T_{K^*}$, given in Eqs.~(\ref{eq:at})-(\ref{eq:ac}) and
Eqs.~(\ref{eq:kss})-(\ref{eq:ksp}), we have already included the
 form-factors.  The form of $f_{\rm c}$ is chosen such that the
 on-shell values of the coupling constants are reproduced.

\subsection{String quark-gluon reaction mechanism}
\label{sec:sqg}
We introduce now the quark-gluon reaction mechanism, based on the work
of Kaidalov~\cite{kaidalov82,kaidalov00}.  It is obvious from the
analysis of the experimental hadron cross section data that the
Reggeon and the Pomeron exchange mechanisms play a crucial role at
high energies, which was nicely demonstrated by Donnachie and
Landshoff~\cite{donnachie87}.  Kaidalov demonstrated that the
quark-gluon mechanism involves the formation of the QCD string between
colored objects formed by high energy collisions and the reaction
cross section could be related with the Regge-slope, $\alpha'$, and
the intercept, $\alpha(0)$, in the Reggeon exchange model.  Hence the
reaction $\gamma p \to K^+\Lambda^*$ can be described by the exchange
of two valence ($u$ and $\bar s$) quarks in the $t-$channel with any
number of gluon exchanges between them. Alternatively, in terms of the
Regge phenomenology, it corresponds to a Reggeon ($R$) exchange, and
thus the scattering amplitude reads,
\begin{equation}
T_{qgR}= \frac{\bar g_{\gamma K R} \bar g_{R N \Lambda^* }}{M_R} 
 (-s/s_0)^{\alpha(t)}F(t)
\label{eq:tqg}
\end{equation}
following the work of Grishina et al.~\cite{grishina06}.  $F(t)$ is a
form-factor which accounts for the compositeness of the external
(incoming and outgoing) hadrons with $t=q^2$, the squared of the four momentum
transfer. We take $F(t) = F_{\gamma K}(t) F_{ p\Lambda^*}(t)$ with
gaussian forms for each of the vertices, i.e. 
$F_\beta(t)=\exp(t/a_\beta^2)$, and then the combined one is  
 also of gaussian type, 
\begin{equation}
F(t)=\exp(t/a^2) \label{eq:ff}
\end{equation}
The constant $s_0$ is taken as the Mandelstam variable $s$ at
threshold ($s_0=(m_K+M_{\Lambda^*})^2$), and it is introduced to fix
the dimensions and to normalize the coupling constants.  On the other
hand, $\alpha(t)=\alpha(0)+\alpha' t$ is the Regge trajectory
associated to the Reggeon quantum numbers.  We can choose any of the
$K-$ or $K^*-$trajectories or considering both; we denote the
different possibilities by $R$ in Eq.~(\ref{eq:tqg}).  On the other
hand, it is customary to fix the coupling constants to those used in
the effective hadron Lagrangian approach.  However, this is not
necessarily true, since the exchanged Reggeon has its own extended
quark-gluon structure, and it does not have to couple to the external
hadrons with the same strength as the virtual exchanged meson does
within the effective Lagrangian model. In addition, the strength of
the couplings will also depend on the election of $s_0$.  In
Eq.~(\ref{eq:tqg}), we use bars ($\bar g's$) to differentiate the
couplings constants from those appearing in the previous
subsection. The parameter $a$ in Eq.~(\ref{eq:ff}), which controls the
$t-$exponential decrease of the form factor and the coupling
constants will be fixed to the experimental data.

The spin structure of the Reggeon exchange comes from the quark
rearrangement process~\cite{kubo99}, for simplicity we do not consider
it in this study and thus, we consider the Regge amplitude independent
of the  incoming and outgoing particle helicities.

\subsection {Hybrid hadron and Reggeon exchange model}
\label{sec:hybrid}

We propose a hybrid mechanism to study the $\gamma p \to \Lambda^*
K^+$ reaction in a wide range of laboratory photon energies. At low
energies, near threshold, we consider the effective Lagrangian model
discussed in Subsect.~\ref{sec:effect}, while for higher photon
energies we assume that the string quark-gluon mechanism is dominant.
We will implement a smooth transition between both reaction mechanisms
for laboratory photon   energies around 2.5~GeV.  The invariant
differential cross section, $d\sigma/dt$ reads
\begin{equation}
\frac{d\sigma}{dt} = \frac{1}{4\pi}\frac{M_N
  M_{\Lambda^*}}{(s-M_N^2)^2} \,
  \Big (\frac{1}{4}\sum_{\lambda, s_p, s_\Lambda} |T|^2 \Big )
\end{equation}
where the invariant Mandelstam variable $t=(k_1-p_1)^2$ varies in the
range $t_-\le t \le t_+$, with $t_{\pm}=m_K^2-2|\vec{k}_1^{\rm \,
cm}|\left ( p_1^0\mp|\vec{p}_1| \right)^{\rm cm}$, with variables
defined, for instance, in the center of mass (cm) frame. It is also of
interest the angular differential cross section in the cm frame, which
is related to $d\sigma/dt$ by
\begin{equation}
\left.\frac{d\sigma}{d\Omega}\right|_{\rm cm} =
\frac{|\vec{k}_1^{\rm \,cm}| |\vec{p}_1^{\rm \,\,cm}|}{\pi}\frac{d\sigma}{dt}
\end{equation}
The sum over polarizations is trivially done for the Regge amplitude,
since we have neglected any spin dependence in that case. In the case
of the  effective Lagrangian approach, it can be easily done thanks to
\begin{equation}
\sum_{\lambda} \epsilon^\mu (k_1, \lambda)\epsilon^{\nu *} (k_1,
  \lambda) = -g^{\mu\nu} + \cdots 
\end{equation}
where the terms $\cdots$ are proportional to $k_1^\mu$ and/or
$k_1^\nu$ and do not contribute because of gauge invariance, and  the
traditional expressions for the sum of Dirac  and 
Rarita-Schwinger spinors. In this latter case, we use
\begin{eqnarray}
\sum_{s_\Lambda} u^\mu(p_2, s_\Lambda)\bar u^\nu(p_2, s_\Lambda) &=& -
  \frac{\Slash{p}_2+M_{\Lambda^*}}{2 M_{\Lambda^*}} P^{\mu\nu}(p_2)\nonumber \\
 P^{\mu\nu}(p_2) &=& \left (g^{\mu\nu}-
  \frac13 \gamma^\mu\gamma^\nu-\frac23\frac{p_2^\mu
  p_2^\nu}{M_{\Lambda^*}^2}+ \frac13\frac{p_2^\mu
  \gamma^\nu-p_2^\nu \gamma^\mu }{M_{\Lambda^*}}\right ) 
\end{eqnarray}
Finally, we get
\begin{equation}
\frac{1}{4}\sum_{\lambda, s_p, s_\Lambda} |T|^2  \, = \,
\frac{1}{16M_NM_{\Lambda^*}} g_{\nu\sigma} {\rm Tr}\left
  ((\Slash{p}_2+M_{\Lambda^*})P_{\rho\mu}(p_2)A^{\mu\nu}(\Slash{k}_2+M_{N}) \gamma^0 (A^{\rho\sigma})^\dagger\gamma^0  \right)
\end{equation}
with 
\begin{equation}
A^{\mu\nu} = \sum_i A^{\mu\nu}_i, \qquad i=s,c,t,v~{\rm or}~v'
\end{equation}
We construct the $T-$matrix as a weighted combination of the two
reaction mechanism contributions,
\begin{eqnarray}
T&=& T({\rm hadron}) \left(1-g(E_\gamma^{\rm LAB})\right) + T({\rm quark})
g(E_\gamma^{\rm LAB}), \nonumber\\
&& T({\rm hadron})  =  T_K+T_{K^*}, \qquad T({\rm quark})=T_{qgR}.
\end{eqnarray}
and for the weighting function $g(E)$, we  use
\begin{equation}
g(E)=\frac{1}{1+exp(-(E-E_0)/\Delta E)}
\label{eq:weight}
\end{equation}
We will fix $E_0$ and $\Delta E$ by comparing with the
experimental data.

\section{SU(6) WT unitary model ($\chi$SU(6)-BSE) and the 
$\Lambda(1520)$ resonance}
\label{sec:su6}

The WT interaction Lagrangian, which is the leading contribution
in the chiral counting, has been the starting point of all SU(3)
chiral unitarity approaches developed in the recent years to study
meson-baryon s-wave scattering. As discussed in the introduction,  SU(6)
spin-flavor symmetry might provide a reasonable framework where to
incorporate baryon decuplet and vector meson nonet degrees of freedom
in the study of hadron processes at low energies.  A
consistent SU(6) extension of the WT SU(3) chiral Lagrangian was
presented in Ref.~\cite{garcia06-su6}, and its generalization to an
arbitrary number of colors and of colors and flavors can be found in
Refs.~\cite{garcia06nc} and \cite{garcia07nc}, respectively.

 Following Ref.~\cite{garcia06-su6}, the building blocks of this
 extension are the \{35\} and \{56\} representations\footnote{We label the
 SU(6) multiplets by a number, which is their dimensionality, enclosed
 between curly brackets.} of SU(6). The first one is the adjoint
 representation of the SU(6) group, and its SU(3) multiplet and SU(2)
 spin content is
\begin{equation}
\{35\}=8_1 + 8_3+1_3
\end{equation} 
where we denote a SU(3) multiplet $\mu$ of spin $J$ by $\mu_{2J+1}$.
Hence, the pseudoscalar meson octet ($K$, $\pi$, $\eta$ and $\bar K$), and
the vector meson nonet ($K^*$, $\rho$, $\omega$, $\bar K^*$ and
$\phi$) belong to the same SU(6) representation,\{35\}.  The lowest mass
baryons belong to the \{56\} representation of SU(6), which is totally
symmetric allowing the baryon to be made of three quarks in
$s-$wave. Its spin-flavor (isospin and hypercharge) content is
determined by the decomposition
\begin{equation}
\{56\}=8_2 + 10_4
\end{equation}
and it accommodates  the spin $1/2^+$ members of the nucleon octet ($N$,
$\Sigma$, $\Lambda$ and $\Xi$) and the spin $3/2^+$  members of the
$\Delta$ decuplet ($\Delta$, $\Sigma^*$, $\Xi^*$ and $\Omega$).

We denote a meson state as $\mathcal{M}=[(\mu_M)_{2J_M+1},I_M,Y_M]$,
where $J_M$, $I_M$, $Y_M$ are the spin, isospin and the hypercharge
quantum numbers of the meson. We use a similar notation for baryon
states $\mathcal{B}$. The meson-baryon states are then expressed in
terms of the SU(6) coupled basis, $\left| \phi; \mu_{2J+1}^\alpha IY
\right\rangle$, as
\begin{eqnarray}
|{\cal M} {\cal B}; JIY \rangle= \sum_
{\mu,\alpha,\phi} 
\left( \begin{array}{cc|c} \mu_M& \mu_B&
   \mu \\ I_MY_M & I_B Y_B &
   IY\end{array}\right) 
 \nonumber \\ 
\times
\left( \begin{array}{cc|c} 35& 56&
   \phi \\ \mu_M J_M & \mu_B J_B &
   \mu J\alpha \end{array} \right)
  \left| \phi; \mu_{2J+1}^\alpha IY \right\rangle 
\,,
  \label{eq:estado}
\end{eqnarray}
where the first and second factors in the linear combination are SU(3)
and SU(6) Clebsh-Gordan coefficients, respectively, and are given in
Refs.~\cite{swart63} and~\cite{carter65}.  In the SU(6) coupled basis,
$\phi$ stands for the SU(6) irreducible representations
($\phi=\{56\}$, $\{70\}$, $\{700\}$ and $\{1134\}$, from the reduction
into irreducible representations of the product $\{35\}\otimes
\{56\}$) and $\alpha$ accounts for possible multiplicity of each of
the $\mu_{2J+1}$ SU(3) multiplets of spin $J$.

The assumption that the s-wave meson-baryon potential, $V$, is a SU(6)
invariant operator implies that $|\phi; \mu^\alpha_{2J+1} I Y\rangle$
coupled states are eigenvectors of the potential and the corresponding
eigenvalues ($V_\phi(s)$), besides the Mandelstam variable $s$, depend
only on the SU(6) representation $\phi$, being thus independent of the
other quantum numbers, $\mu, \alpha, J, I, Y$.  Hence, the matrix
elements of the potential can be written as
\begin{equation}
V^{JIY}_{\mathcal{MB,M'B'}}(s)=\langle
\mathcal{M',B'};JIY|V|\mathcal{M,B};JIY\rangle =\sum_\phi V_\phi(s)
\mathcal{P}^{\phi,JIY}_{\mathcal{MB,M'B'}} \label{eq:su6a}
\end{equation}
with the projection operators given by
\begin{eqnarray}
&&
{\cal P}_{{\cal M}{\cal B}, {\cal M}^\prime 
{\cal B}^\prime }^{\phi,JIY}
= \sum_{\mu,\alpha} 
\left( \begin{array}{cc|c} 35& 56&
   \phi \\ \mu_M J_M & \mu_B J_B &
   \mu J \alpha \end{array}\right)\nonumber\\
&\times& 
\left( \begin{array}{cc|c}
 \mu_M& \mu_B& \mu \\ 
  I_M Y_M & I_B Y_B & I Y\end{array}\right)
\left( \begin{array}{cc|c} \mu^\prime_{M^\prime}& \mu_{B^\prime}^\prime&
   \mu \\ I^\prime_{M^\prime}Y^\prime_{M^\prime} &
 I^\prime_{B^\prime} Y^\prime_{B^\prime} &
   IY\end{array}\right) 
\nonumber\\
&\times&
\left( \begin{array}{cc|c} 35& 56&
   \phi \\ \mu'_{M'} J'_{M'} & \mu'_{B'} J'_{B'} &
   \mu J \alpha \end{array}\right). \label{eq:su6b}
 \phantom{hhhhhhhhhhhhh}
\end{eqnarray}

We are now in a position to extract the four SU(6) eigenvalues by
relating them to those of the SU(3) matrix elements of the WT
interaction.  The mesons of the pion octet and the baryons of the
nucleon octet interact through the WT Lagrangian, being the
corresponding potential~\cite{kaiser95,oset98,oller01,nieves01}
\begin{eqnarray}
V^{IY}_{ij}(\sqrt{s})& = &
D^{IY}_{ij} \frac{\sqrt{s}-M}{2\,f^2} 
\label{eq:su3a}
\end{eqnarray}
where the indices $i$ and $j$ identify the final and initial meson-baryon
pair (quantum numbers $I^\prime_{M^\prime} Y^\prime_{M^\prime}
I^\prime_{B^\prime} Y^\prime_{B^\prime}$ and $I_{M}
Y_{M} I_{B} Y_{B}$, respectively) and
\begin{eqnarray}
D^{IY}_{ij} &=& \sum_{\mu,\gamma,\gamma^\prime}
\lambda_{\mu_\gamma \to \mu_{\gamma^\prime}} \left (
\begin{array}{cc|c} 8 & 8 & \mu_\gamma \\ 
I_M Y_M & I_B Y_B & IY
\end{array}
\right)
\nonumber \\
&\times& \left (
\begin{array}{cc|c} 8 & 8 & \mu_{\gamma^\prime} \\ 
I^\prime_{M^\prime} Y^\prime_{M^\prime} & I^\prime_{B^\prime} Y^\prime_{B^\prime} &
  IY\end{array}\right) \label{eq:su3b}
\end{eqnarray}
Here, $M$ is the baryon mass, $f\simeq 93\,$MeV the pion weak decay
constant, $\mu$ runs over the $27$, $10$, $10^*$, $8$ and $1$ SU(3)
representations and $\gamma,\gamma^\prime$ are used to account for the
two octets ($8_s$ and $8_a$) that appear in the reduction of
$8\otimes 8$. The SU(3) eigenvalues $\lambda$'s, are
$\lambda_{27}= 2$, $\lambda_{8_s}= \lambda_{8_a}=-3$,
$\lambda_{1}=-6$, $\lambda_{10} =
\lambda_{10^*}=\lambda_{8_s\leftrightarrow 8_a}
=0$~\cite{garcia04}. Note how chiral symmetry determines all
eigenvalues, which otherwise will be totally independent and unknown
for a generic SU(3) symmetric theory. 

To deduce the SU(6) eigenvalues,
we study the reduction of the SU(6) matrix elements of
Eqs.~(\ref{eq:su6a})-(\ref{eq:su6b}) when only the pion and nucleon
octets are considered. It is clear that not all SU(3) invariant
interactions in the $(8_1)$meson--$(8_2)$baryon sector can be extended
to a SU(6) invariant interaction. Remarkably, the seven couplings
($\lambda$'s) in the WT interaction turn out to be consistent with
SU(6) and moreover, the extension is unique. Indeed, we find that by taking
\begin{equation}
V_{\phi}(s) = \bar\lambda_{\phi}
\frac{\sqrt{s}-M}{2\,f^2}\,,
\label{eq:vsu6} 
\end{equation}
with the coefficients $\bar \lambda$'s given by
\begin{equation} 
\bar\lambda_{56}=-12,\quad \bar\lambda_{70}=-18, \quad 
\bar\lambda_{700}=6, \quad \bar\lambda_{1134}=-2, \label{eq:lssu6}
\end{equation} 
the SU(3) matrix elements described above 
(Eqs.~(\ref{eq:su3a})-(\ref{eq:su3b}))  are completely
reproduced. As it is discussed in Ref.~\cite{garcia06-su6}, the
underlying chiral symmetry of the WT Lagrangian has made possible the
spin symmetry extension presented here.
 
 Next we consider explicit SU(6) breaking effects due to the use of
physical (experimental) hadron masses and  meson decay constants.
Hence in Eq.~(\ref{eq:vsu6}), we make the replacement 
\begin{equation}
\frac{\sqrt{s}-M}{2f^2} \rightarrow \frac{2\sqrt{s}-M_i-M_j}{4 f_i f_j}
\end{equation}

To describe the dynamics of resonances one needs to have exact elastic
unitarity in coupled channels. For that purpose, we solve the coupled
channel BSE and use the SU(6) potential defined above to construct its
interaction kernel. In a given $JIY$ sector, the solution for the
coupled channel $s-$wave scattering amplitude, $T^{J}_{IY}$, satisfies
exact unitarity in coupled channels. In the so called {\it on-shell}
scheme~\cite{garcia04}, and normalized as the $t$ matrix defined in
Eq.~(33) of the first entry of Ref~\protect\cite{nieves01}, it is
given by
\begin{equation}
T_{IY}^J(s)=\frac{1}{1-V_{IY}^J(s)J_{IY}^J(s)}V_{IY}^J(s) \label{eq:t-matrix}
\end{equation}
where $J_{IY}^J(s)$ is a diagonal function in the coupled channel
space. Suppressing the indices, it is written for each channel as
\begin{eqnarray}
J(s)&=&\frac{(\sqrt{s}+M)^2-m^2}{2\sqrt{s}}J_0(s)\\
J_0(s)&=& i\int \frac{d^4k}{(2\pi)^4} \frac{1}{(P-k)^2 -M^2+i\epsilon}
\,\frac{1}{k^2-m^2+i\epsilon}
\end{eqnarray}
where $M$ and $m$ are the masses of the baryon and meson corresponding
to the channel and $P^\mu$ is the total meson-baryon four momentum
($P^2=s$). On the other hand, $J_0(s)$ involves a logarithmic
divergence which needs to be subtracted,
\begin{equation}
J_0(s)=\bar J_0(s) + J_0(s=(M+m)^2)
\end{equation}
where the finite $\bar J_0(s)$ function can be found in Eq.~(A9) of
the first entry of Ref~\protect\cite{nieves01}). It induces the
unitarity right hand cut of the amplitude. Besides, the constant 
$J_0(s=(M+m)^2)$ hides the logarithmic divergence. It is renormalized
by requiring 
\begin{equation}
J_0(s=\mu_0^2) = 0 \label{eq:rs}
\end{equation}
at a certain scale $\mu_0$. This defines our RS. A suitable choice is
to take $\mu_0$ independent of $J$ and set it uniformly within a given $IY$
sector as $\sqrt{m_{\text{th}}^2+M^2_{\text{th}}}$, where
$(m_{\text{th}}+M_{\text{th}})^2$ gives the smallest threshold among
all channels involved in a $IY$ sector~\cite{LK02}.

Particularly for the $\Lambda(1520)$ state, with quantum numbers
$I=0$, $Y=0$ and $J^\pi=3/2^-$, we have a total of 9 channels in the
SU(6) scheme: $\pi \Sigma^* $, $ \bar K^* N  
  $, $ \omega \Lambda $, $ \rho\Sigma   $, $ K\Xi^* $, $ \phi \Lambda
$, $\rho \Sigma^* $, $  K^* \Xi  $ and $  K^* \Xi^*$. The potential
thus reads,
\begin{equation}
V^{(3/2,0,0)}= D^{(3/2,0,0)} \frac{2\sqrt{s}-M_i-M_j}{4 f_i f_j}
\end{equation}
where the symmetric coupled channel matrix $D^{(3/2,0,0)}$ is obtained
  from Eqs.~(\ref{eq:su6a})-(\ref{eq:su6b}) and 
(\ref{eq:vsu6})-(\ref{eq:lssu6}).  It
  is given by
\begin{eqnarray} 
&& D^{(3/2,0,0)} =  \\
&& \left( \begin{matrix}
 -4 & -\sqrt{2} & 0        & -4/\sqrt{3} & -\sqrt{6} & 0 & -\sqrt{80/3} & \sqrt{2}    &-\sqrt{10}  \vspace*{.2cm}\cr
    &  0        & \sqrt{6} & \sqrt{2/3}  & 0         & 0 &  \sqrt{10/3} &  0          & 0          \vspace*{.2cm}\cr
    &           & 0        & 2           &  \sqrt{2} & 0 &  \sqrt{20}   & -\sqrt{2/3} & \sqrt{10/3}\vspace*{.2cm}\cr
    &           &          & -8/3        & -\sqrt{2} & 0 & -\sqrt{20/9} &  \sqrt{2/3} &-\sqrt{10/3}\vspace*{.2cm}\cr
    &           &          &             & -3        &-2 & -\sqrt{10}   &  0          & -\sqrt{15} \vspace*{.2cm}\cr
    &           &          &             &           & 2 &  0           & \sqrt{16/3} & \sqrt{20/3}\vspace*{.2cm}\cr
    &           &          &             &           &   & -16/3        & \sqrt{10/3} &-\sqrt{50/3}\vspace*{.2cm}\cr
    &           &          &             &           &   &              & -4/3        & \sqrt{80/9}\vspace*{.2cm}\cr
    &           &          &             &           &   &              &             & -11/3      \vspace*{.2cm}\cr
\end{matrix} \right) \quad \begin{matrix} \pi \Sigma^*  \vspace*{.2cm}\cr \bar K^* N  
   \vspace*{.2cm}\cr \omega \Lambda  \vspace*{.2cm}\cr 
\rho\Sigma    \vspace*{.2cm}\cr K\Xi^*  \vspace*{.2cm}\cr \phi \Lambda  \vspace*{.2cm}\cr
  \rho \Sigma^*  \vspace*{.2cm}\cr  K^* \Xi   \vspace*{.2cm}\cr  K^* \Xi^*    \vspace*{.2cm} \end{matrix} 
\nonumber
\end{eqnarray} 
Note we assume an ideal mixing in the vector meson sector, i.e.,
$\omega=\frac{1}{\sqrt 2}\left ( u\bar u + d\bar d\right)$ and $\phi=s
\bar s$, which induces the use of some linear combinations of the
isoscalar SU(3) vector meson mathematical states\footnote{In the
vector meson sector, Carter's SU(6)--multiplet coupling
factors~\cite{carter65} are consistent with the election of
$\phi_1=\frac{1}{\sqrt 3}\left ( u\bar u + d\bar d + s\bar s\right)$
and $\phi_8=-\frac{1}{\sqrt 6}\left ( u\bar u + d\bar d -2 s\bar
s\right)$ quark wave functions for the SU(3) singlet and isospin
singlet of the SU(3) octet, respectively. The relative minus sign is
absent in the pseudoscalar meson sector.}.

 We include an explicit breaking of the
 SU(6) symmetry through the use of the experimental masses and the
 meson decay constants: $f_\pi=92.4$ MeV~\cite{pdg}, $f_K=113.0$
 MeV~\cite{pdg}, $f_\rho=f_{K^*}= 153 $ MeV (from
 $\Gamma(\rho\to e^+e^-)$, $\Gamma(\tau\to \rho \nu_\tau)$, and
 $\Gamma(\tau\to K^* \nu_\tau)$), $f_\phi =163$ MeV (from
 $\Gamma(\phi\to e^+e^-)$) and $f_\omega\approx f_\rho$.

\section{Results and discussion}
\label{sec:res}

We use $M_{\Lambda^*}=1519.5 $ MeV and charged nucleon, kaon and $K^*$
  masses from the PDG~\cite{pdg}. Besides, we use $eg_{\gamma K
  K^*}=0.23$ and $g_{K N \Lambda^*}=10.5$ from the $K^{*+} \to K^+
  \gamma$ radiative\footnote{From the Lagrangian of
  Eq.~(\ref{eq:k*rad}), we obtain
\begin{equation}
\Gamma = \frac{1}{96\pi} \left(eg_{\gamma K K^*}\right)^2
m_{K^*}\left(1-\frac{m_K^2}{m_{K^*}^2}\right)^3 \approx 50.3 ~\,{\rm
  KeV}~\cite{pdg} 
\end{equation}
}, and the 
$\Lambda^*$ decay widths\footnote{From the Lagrangian of
  Eq.~(\ref{eq:knl}), we obtain for the $\Lambda^* \to p K^-$ decay width
\begin{equation}
\Gamma = \frac{1}{12\pi}\frac{g_{K N \Lambda^*}^2}{m_K^2} |\vec{p}_{cm}|^3
\frac{M_{\Lambda^*}-p^0_{cm}-M_N}{M_{\Lambda^*}} \approx \frac12
0.45\times 15.6 ~{\rm MeV}~\cite{pdg}. 
\end{equation}
where $p^\mu_{cm}$ is the antikaon four momentum in the $\Lambda^*$
rest frame.}, respectively.  Thus the only unknown parameters in the
effective Lagrangian approach are $g_{K^* N \Lambda^*}$ and the cutoff
$\Lambda$ entering in the hadron form-factors of
Eq.~(\ref{eq:cut-had}).

\begin{figure}[t]
\begin{center}
  \resizebox{150mm}{!}{\includegraphics{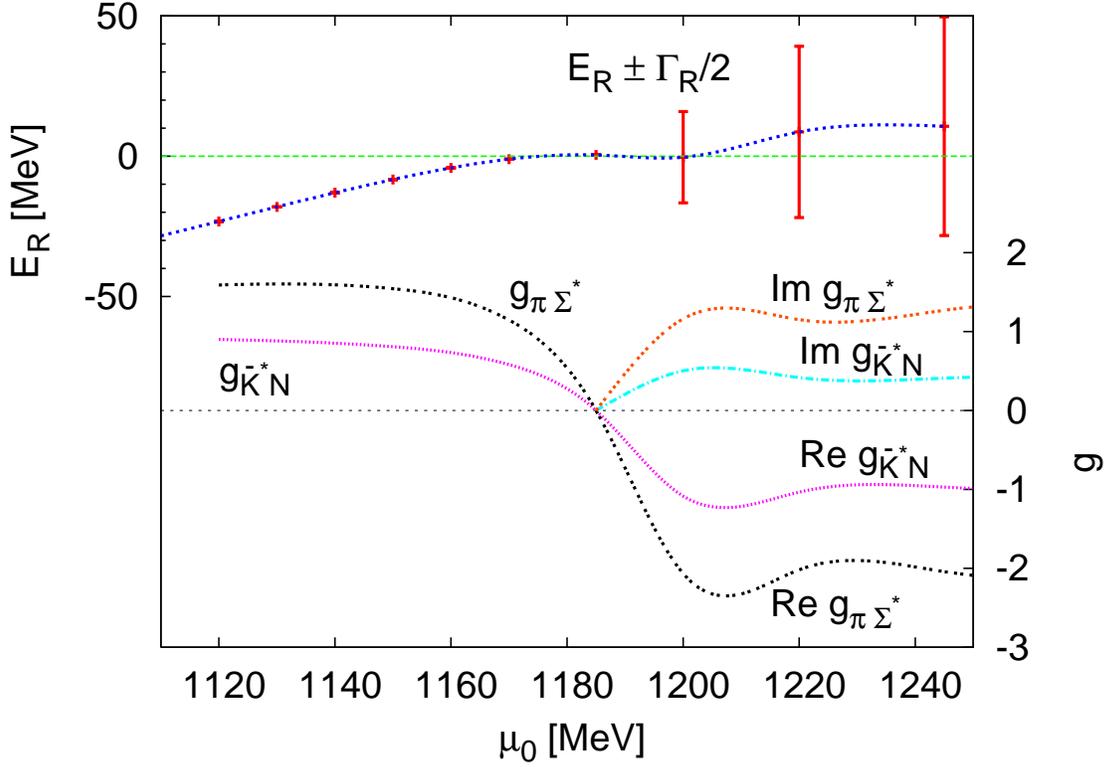}}
  \caption{{\small $\chi$SU(6)-BSE predictions for the $\Lambda^*$
  resonance pole position ($M_R=m_\pi+M_{\Sigma^*}+E_R$,  $\Gamma_R$)
  and couplings (Eq.~(\ref{eq:tpole})) to the $\pi\Sigma^*$ and $\bar
  K^* N$ channels as function of the renormalization scale $\mu_0$.}
  }
  \label{fig:gigj}
\end {center}
\end{figure}
\subsection{The coupling $g_{K^*N\Lambda^*}$}

\label{sec:gknl}
We study here the dynamics of $\Lambda^*$ resonance within the SU(6) model
presented in Sect.~\ref{sec:su6} Without including the vector mesons
in the model, there exist only two channels which contribute to the
formation of the $\Lambda^*$ state, which are $\pi$-$\Sigma^*$ and
$K$-$\Xi^*$.  Since the mass of $\Lambda^*$ resonance is very close to
the threshold energy of the $\pi$-$\Sigma^*$ channel ($\approx 1520-1525$
MeV), this channel dominates the dynamics of $\Lambda^*$.  When the
vector meson degrees of freedom are taken into account, there are
appear seven additional channels.  However, the dynamics of the
$\Lambda^*$ resonance is almost unaffected by the heavier ones.  Thus
in very good approximation, we have considered the $4\times 4$
sub-matrix of $D^{(3/2,0,0)}$ constructed out the first four rows and
columns ($\pi$-$\Sigma^*$, $\bar K^*$-$N$, $\omega$-$\Lambda$ and
$\rho$-$\Sigma$).  We solve the coupled channel BSE and renormalize
the amplitudes as described above in 
Eqs.~(\ref{eq:t-matrix})--(\ref{eq:rs}). 
We look for complex poles of the $T-$matrix in the
Second Riemann Sheet (SRS), determined by continuity to the First
Riemann Sheet (FRS)~\cite{nieves01}. In a given $JIY$ sector, physical
resonances appear in the SRS of all matrix elements of $T(s)$  
(Eq.~(\protect\ref{eq:t-matrix})), in the coupled channel space,
differing only on the value of the residue at the pole.  The pole
position determines the mass and width of the resonance, while the
different residues for each meson-baryon channel give the respective
couplings and branching ratios (see section II.D of the
second entry of Ref.~\cite{nieves01}). Let us consider $s_R = M^2_R -
{\rm i}~ M_R \Gamma_R$ a pole in the SRS of the
coupled channel scattering matrix $T(s)$. Then, around the pole, it
can be approximated by
\begin{eqnarray}
\left[T(s)\right]_{ij} & \approx & 2 M_R \frac{g_i g_j}{s-s_R},
\label{eq:tpole}
\end{eqnarray}
where $g_ig_j$ is the residue matrix. The complex
vector $g_i$ determines the coupling of the resonance to the different
final states, which are well and unambiguously defined even if the
corresponding channels are closed in the decay of the resonance.

As discussed above, after Eq.~(\ref{eq:rs}), to describe the $I=0,
Y=0$ sector, our standard choice would be
$\mu_{0}=\sqrt{m_\pi^2+M_\Sigma^2}\approx 1.2$ GeV, for which we find a
pole with $M_R=1528$ MeV and $\Gamma_R=42$ MeV. This is a remarkable
result, since the SU(6) model predicts the existence of the
$\Lambda^*$ resonance.  Within the model, it appears slightly above
the $\pi\Sigma^*$ threshold and that originates a non-vanishing
width, since the $\pi\Sigma^*$ channel is open. Experimentally, the
mass of the resonance is 1519.5 MeV, around 3 MeV below the $\pi
\Sigma^*$ threshold, and its width is around 15.6 MeV (it decays
predominantly into d-wave $\pi\Sigma$ and $\bar K N$ pairs and the
three body mode $\Lambda \pi \pi$, none of them considered in the
SU(6) model). To better describe the dynamics of the resonance, we
have varied the renormalization scale in the vicinity of 1.2
GeV. Results are displayed in Fig.~\ref{fig:gigj}. For values of
$\mu_0$ below 1185 MeV, the pole appears in the FRS and therefore it
should be interpreted as $\pi \Sigma^*$ bound state. Within our model it
would be stable (zero width), since none of the allowed decay modes
($\pi\Sigma$, $\bar K N$ and $\Lambda \pi\pi$) are  included in our
two body s-wave approach. Scales in the range 1160-1185 MeV provide 
masses around 1520 MeV. In the figure we also show the couplings
(Eq.~(\ref{eq:tpole})) of the resonance $\Lambda^*$ to the $\pi\Sigma^*$
 and $\bar K^* N$ channels ($g_{\pi \Sigma^*}$ and  $g_{\bar K^* N}$,
 respectively). We would like to make three remarks. 
\begin{itemize}
\item Both couplings are determined up to an overall minus sign.
\item The coupling $g_{K^*N\Lambda^*}$ (defined in
  Eq.~(\ref{eq:swaveK*N})) is given by $g_{\bar K^* N}/\sqrt 2$, where
  the $\sqrt{2}$ factor comes trivially from the projection of the $p
  K^{*-}$ state into isospin zero.
\item Both couplings vanish when the $\Lambda^*$ resonance
is placed just at the $\pi \Sigma^*$ threshold. It is to say, when the
resonance can be interpreted as $\pi \Sigma^*$ state bound with zero energy.
\end{itemize}
The latter remark has important phenomenological repercussions, since
the actual position of the $\Lambda^*$ is quite close to the $\pi
\Sigma^*$ threshold and it would imply that $g_{K^*N\Lambda^*}$
coupling should be much smaller (we read off from the figure a value
around $0.75/\sqrt{2}$) than the values used in the phenomenological
analysis of Refs.~\cite{nam05,titov05}. It is easy to understand
this behavior of the coupling constants. Let us start studying the
simple case of the elastic scattering of a meson of mass $m$ off a
baryon of mass $M$. The s-wave scattering amplitude close to threshold
can be then approximated by
\begin{equation}
f(p) = \frac{e^{2\,{\rm i}\delta(p)}-1}{2\,{\rm i}
p}\approx \frac{1}{-\frac{1}{\alpha}+\frac12 r_0^2 p^2+...-ip}, \quad
p^2=2\mu E, ~ s\approx s_{\rm th}+2(m+M)E 
\end{equation}
 with $\alpha$ and $r_0$ the scattering length and effective range,
respectively. Besides, $s_{\rm th}= (m+M)^2$ and $\mu$ is the reduced
mass.  If there exist a bound state at $s=s_B$ very close to 
threshold\footnote{It is to say, the scattering amplitude has a 
  pole for a real value of $s=s_B \le s_{\rm th}$. In
  Fig.~\ref{fig:gigj}, it would correspond to $E_R \le 0$.}, one
can drop the scattering range term and the scattering length can be
approximated by $\alpha \sim \frac{1}{\sqrt{2 \mu |E_B|}}$, where
$E_B(<0)=(s_B-s_{\rm th})/2\sqrt{s_{\rm th}}$ is the binding energy of
the bound state. In this situation for both above ($E>0$) and below
($E<0$) threshold, but close to it, $f$ can be written as
\begin{equation}
f(p) \approx \frac{-1}{\sqrt{2\mu}}
\frac{\sqrt{-E}+ \sqrt{-E_B}}{E-E_B}, 
\end{equation}
and therefore the residue of the $f$ at the pole is given by
$-\sqrt{2|E_B|/\mu}$.  Since the scattering amplitude and the $T$-matrix
close to threshold are related by $f \approx -\frac{M}{4\pi (m+M)}T$, from
Eq.~(\ref{eq:tpole}) we obtain 
\begin{equation}
g^2=4 \pi \frac{m}{\mu}\sqrt{\frac{2|E_B|}{\mu}} \label{eq:g2}
\end{equation}
Therefore the square of the coupling of the bound state to the channel
scales like the square root of the binding energy.  The same result
can be found just by looking directly for poles of $T$
(Eq.~(\ref{eq:t-matrix})) in the FRS. Indeed, following
Ref.~\cite{garcia04}, the position of the pole, $s_B$, is such that
the dimensionless function
\begin{equation}
\beta(s) = \frac{2f^2}{J(s)(\sqrt{s}-M)},
\end{equation}
at $s=s_B$, becomes\footnote{In the case of only one channel, $D$ is a
  1$\times$1 matrix, it is to say a real number.} $D$. In addition,
\begin{equation}
g^2= D^2 \frac{\sqrt{s_B}-M}{2f^2} \frac{1}{\beta'(s_B)}
\end{equation}
and it reduces to Eq.~(\ref{eq:g2}) when $s_B$ is close to threshold.
The behaviour of $g^2$ with $|E_B|$ follows from the behaviour of
$dJ/ds|_{s_B}$ when $s_B$ is close to $s_{\rm th}$, where it diverges
like $1/\sqrt{s_{\rm th}-s_B}\approx (2(m+M))^{-1/2}/\sqrt{|E_B|}$,
and therefore $\beta'(s_B)$ does also diverge in that limit.
 
Similar conclusions can be drawn in the general case, when coupled
channels are considered. However, in that case, $D$, $\beta$ and $J$
are matrices and some subtleties appear. Nevertheless, the behavior of
the coupling of the bound state to the different channels can be
analytically worked out.  We find that, if there exists a bound state
very close to the smallest of the thresholds (let us take an ordering
such that this channel is the first one), the sum of the squares of
the couplings of this bound state to the different channels vanishes
as the binding energy approaches to zero.  Indeed, it can be proved
that this sum scales again as $\sqrt{|E_B|}$, i.e.
\begin{equation}
\sum_i g_i^2  = \frac{1}{c^2 \frac{1}{4 \pi
  \frac{m}{\mu}\sqrt{\frac{2|E_B|}{\mu}}}+\cdots}
\end{equation}
where $c$ is a coefficient related with the projection of the
resonance wave function into the first channel, $m$ and $M$ are the
masses of the meson and baryon in this channel,
$E_B=(s_B-(m+M)^2)/2(m+M)$, and the dots stand for some contributions
which remain finite in the $|E_B|\to 0$ limit. Note that the sum over
$i$ in the above equation runs over all channels, and since all
couplings are real, it implies that all terms in the sum must vanish.
For $|E_B| \approx 2.5 $ MeV, we find $\sum_i g^2_i \approx 2.75$,
which is largely saturated by $g_{\pi\Sigma^*}$, leaving little room
for $g_{\bar K^* N}$.

Of course, if one is working with an unique channel, $c=1$, we recover
Eq.~(\ref{eq:g2}). However, if the resonance does not couple to the first
channel, $c=0$ and $g_1=0$, then we cannot conclude anything about the
rest of the couplings.

The results of Fig.~\ref{fig:gigj} favor values of the coupling
constant $g_{K^*N\Lambda^*}$ around 0.5 or smaller, which, following the
discussion below Eq.~(\ref{eq:ksp}), will correspond to a value of
$g'_{K^*N\Lambda^*}$ of about 0.75 and it constitutes one of the major
results of this work. This value is significantly smaller than the
values of around 10 used in the phenomenological analysis of Nam et
al.~\cite{nam05} and Titov et al.~\cite{titov05}. From the
discussion above and taking into account the proximity of the
$\Lambda^*$ to the $\pi \Sigma^*$ threshold, we have compelling
reasons to expect such a small value for this coupling constant.

Hyodo et al.~\cite{hyodo06} find $g'_{K^*N\Lambda^*}\sim 1.5/\sqrt2 $
within an extended SU(3) chiral unitary model\footnote{The $\sqrt{2}$
  factor is implemented here because the value 1.5 quoted in
  Ref.~\cite{hyodo06} refers to the $ K^*N\Lambda^*$ coupling in
  isospin basis ($I=0$).}, which accounts for the baryons of the
decuplet and for some d-wave contributions, and where an effective
Lagrangian is used to include the $\bar K^*$ degrees of freedom.  Our
approach, instead provides directly the coupling of the $\Lambda^*$ to
$\bar K^* N$ because $\bar K^* N$, together with other vector-meson
baryon states, are part of the basis of the coupled channels. These
latter states are not considered in the coupled channel approach of
Ref.~\cite{hyodo06}.  Nevertheless, it is reassuring that the value
quoted in ~\cite{hyodo06} for $g'_{K^*N\Lambda^*}\sim
1.5/\sqrt{2}\approx 1.1 $ is similar to ours ($\approx 0.75$), and in
anycase it is much smaller than that used in the previous
phenomenological studies of the $\gamma p \to K^+ \Lambda^* $
reaction.

From the findings of this subsection, we conclude that the $K^*$
mechanism is much smaller than the $K$ one. We drop hereafter
completely the $K^*$ contribution in the effective Lagrangian
approach.

\subsection{$\gamma p \to K^+ \Lambda^* $ cross section}
\label{sec:res-sigma}
\begin{figure}[t]
\begin{center}
 \resizebox{120mm}{!}{\includegraphics{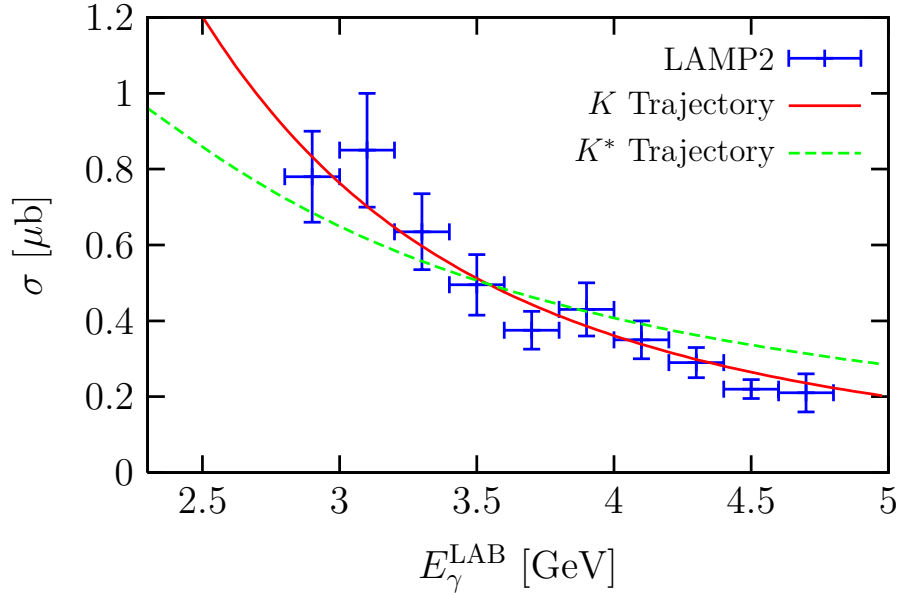}}
 \caption{{\small $\gamma p \rightarrow K^+ \Lambda^*$ total cross
  section (in units of $\mu$b) from the quark-gluon reaction mechanism, as a
  function of the photon energy in the LAB frame.  The solid (dashed) curve
  has been obtained with the $K-$($K^*-$)Regge trajectory. The cutoff
  parameter $a$ in the gaussian form-factor of Eq.~(\ref{eq:ff}) is set to 1
  GeV.  The experimental data are taken from Ref. \cite{barber80}.}  }
  \label{fig:figkks}
\end {center}
\end{figure}
\begin{figure}[t]
\begin{center}
  \resizebox{120mm}{!}{\includegraphics{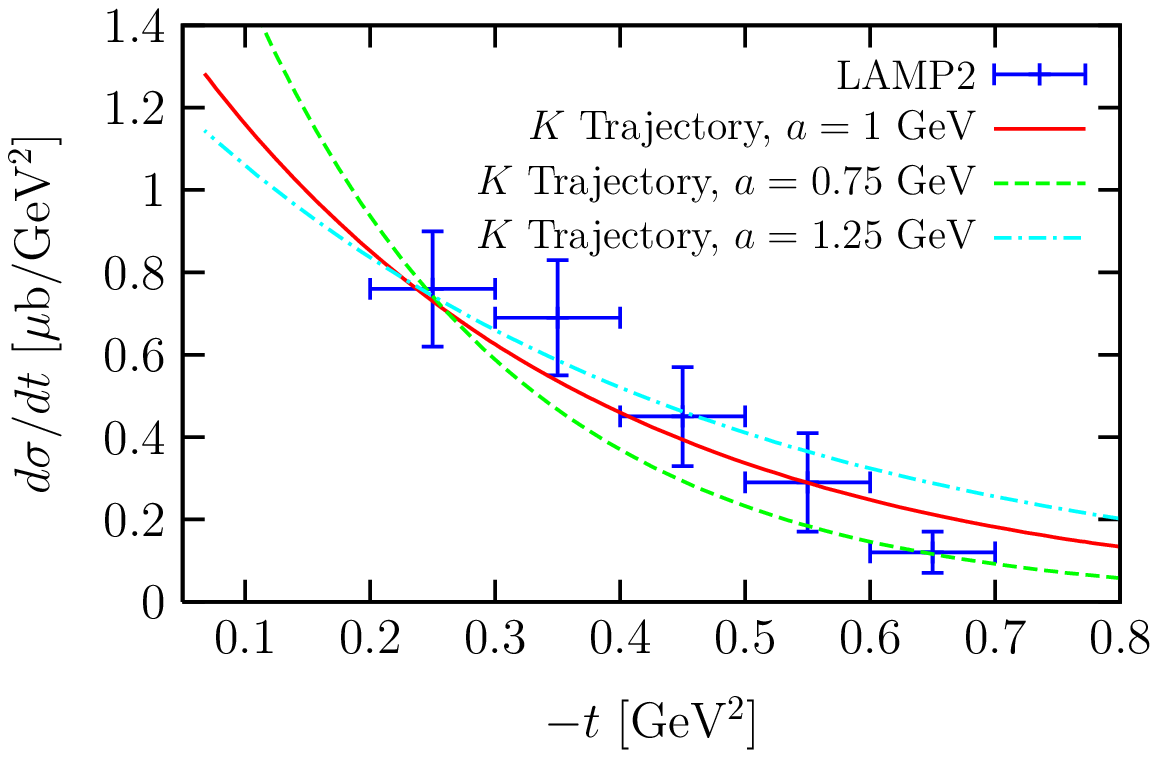}}
  \caption{\small $\gamma p \rightarrow K^+ \Lambda^*$ differential
    cross section ($d\sigma/dt$) from the $K-$Reggeon exchange
    mechanism, as a function of $(-t)$ for $E_\gamma^{\rm LAB}=3.8$
    GeV.  We use three values for the cutoff parameter $a$ in the
    gaussian form-factor of Eq.~(\ref{eq:ff}): $a=1$ GeV, $a=0.75$ GeV
    and $a=1.25$ GeV, which results correspond to the solid, dashed
    and dash-dotted curves, respectively.  In the latter two cases,
    the results have been scaled up and down by factors 3/2 and 0.85,
    respectively. The experimental data-points, taken from
    \cite{barber80}, stand for the $K^+\Lambda^*$ photo-production
    differential cross section $d\sigma/dt$ averaged over the
    incident LAB photon energy range $ 2.8~{\rm GeV} \le E_\gamma^{\rm
    LAB} \le 4.8 ~{\rm GeV}$ of the Daresbury experiment.}
  \label{fig:figt}
\end {center}
\end{figure}
We use a hybrid approach, and assume that the quark-gluon string
mechanism dominates the $\gamma p \to K^+ \Lambda^* $ reaction at
photon energies well above threshold, where the experimental data
exist.  The quark-gluon string model accounts, in principle, for both
the $K$ and the $K^*$ Reggeon exchange processes.  For these
trajectories, we use
\begin{eqnarray}
\alpha_K(t)&=&-0.20+0.8\,t/{\rm GeV}^2 \\
\alpha_{K^*}(t)&=&\phantom{-}0.36+0.8\,t/{\rm GeV}^2, 
\end{eqnarray}
where we have taken slopes around 0.8 [GeV$^{-2}$] that provide, as
we will show, a reasonable description of the $d\sigma/dt$ data, and are
not far from the value of around 0.9 used in
Ref.~\cite{titov05}\footnote{Note, however, that the $K^*-$trajectory
  used in this reference does not lead to $\alpha_{K^*}(m_{K^*}^2)=1$.
  Indeed, it is much closer to our $K-$trajectory than to the
  $K^*-$one, in the $t-$range $[-1:0]$ GeV$^2$ of interest in this
  work. } or of 0.93 deduced from the
$\rho-$trajectory~\cite{donnachie92}. The intercepts, $\alpha(0)$, 
are determined by
requiring $\alpha_K(m_{K}^2)=0$ and $\alpha_{K^*}(m_{K^*}^2)=1$.
\begin{figure}[t]
\begin{center}
  \resizebox{120mm}{!}{\includegraphics{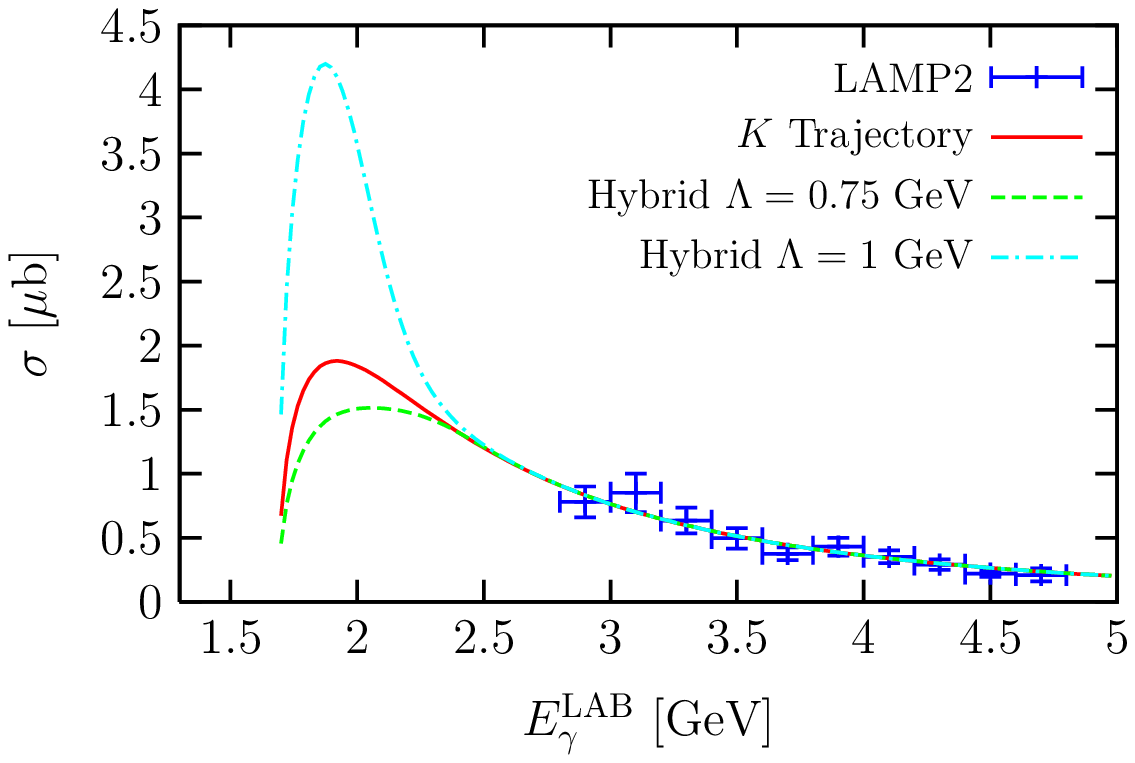}}
  \caption{\small Predictions from different models and
data~\cite{barber80} for the $\gamma p \rightarrow K^+ \Lambda^*$
total cross section as a function of the LAB photon energy.  Results
from the quark-gluon mechanism ($a=1$ GeV) are shown by the solid
curve. The other two curves display the hybrid hadron/$K-$Reggeon
exchange model results (see Subsect.~\ref{sec:hybrid}) for two
different values of the cutoff $\Lambda$ entering in the hadron
form-factors of Eq.~(\ref{eq:cut-had}). The dashed and dash-dotted
lines stand for the results obtained with $\Lambda=0.75$ GeV and
$\Lambda=1$ GeV, respectively. In the latter (former) case  we use 
$E_0=2.0 (2.3)$ GeV, while  the parameter $\Delta E$ is set in both
cases to 0.1 GeV. }
  \label{fig:figtl}
\end {center}
\end{figure}

We show in Fig.~\ref{fig:figkks} the energy dependence of the photon
induced $\Lambda^*$ production cross section with both Regge
trajectories.  We use a cutoff $a$ in Eq.~(\ref{eq:ff}) of 1 GeV,
which is of a natural size in hadron physics. To describe the overall
normalization of the data, we take $\bar g_{\gamma K K} \bar g_{K N
  \Lambda^*} = eg_{K N \Lambda^*} \times 0.12$ and $\bar g_{\gamma K
  K^*} \bar g_{K^* N \Lambda^* } = eg_{\gamma K K^*} g_{K^* N \Lambda^*}
\times 4.24$. We see that the $K$ Regge trajectory provides a better
description of the data than the $K^*$ one in the energy range studied
here. Of course, this conclusion is affected by the choice of the
cutoff parameter $a$ in the gaussian form-factor of Eq.~(\ref{eq:ff}).
The larger $a$, the better description of data is obtained with the
$K^*-$trajectory, since the slope (in absolute value) of the cross
section with respect $E_{\gamma}^{\rm LAB}$ increases.  Nevertheless,
the description obtained from the $K-$trajectory is always better. To
find a more or less comparable description from both trajectories we
need to use values of $a$ above 10 GeV. For those values of $a$ and
the energies explored in this work, the form-factor $F$ is in practice
1, which is somehow unrealistic. We remind here also that to make both
the $K-$ and $K^*-$Regge contributions similar in size, we should
re-scale one over the other by a factor of the order 1200 [
$(4.24/0.12)^2$].

Thus, here again we conclude that the $K$ Reggeon mechanism is more favored
by data than the $K^*$ Reggeon one, which will be neglected in what follows.

\begin{figure}[t]
\begin{center}
  \resizebox{120mm}{!}{\includegraphics{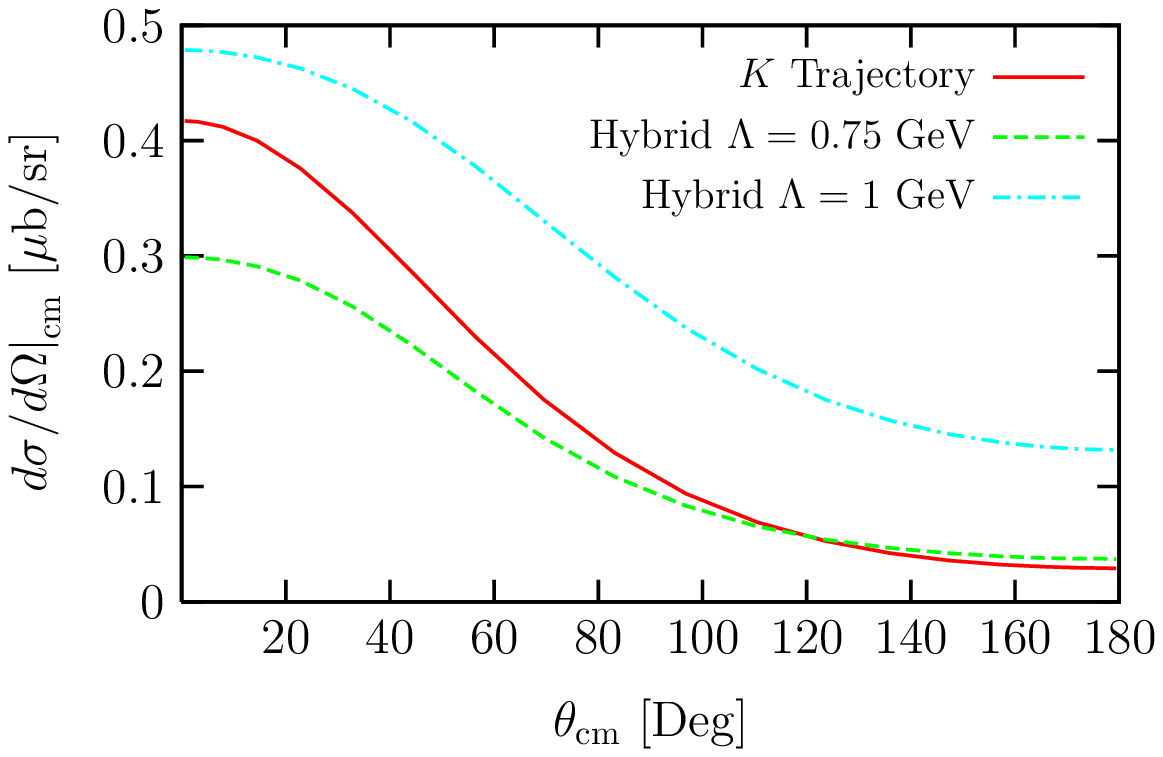}}
  \caption{\small Center of mass differential cross section
predicted by the three theoretical models of Fig.~\ref{fig:figtl} for
$E_\gamma^{\rm LAB}=2$ GeV.}
  \label{fig:figad}
\end {center}
\end{figure}
Next, we pay attention to $d\sigma/dt$ and test the dependence of the
Regge results on the cutoff
parameter $a$. This
differential cross section, averaged over the incident LAB photon
energy range $ 2.8~{\rm GeV} \le E_\gamma^{\rm LAB} \le 4.8 ~{\rm
  GeV}$, has been measured in the Daresbury
experiment~\cite{barber80}.  In Fig.~\ref{fig:figt}, we compare these
measurements with our $K-$Reggeon exchange results for three values of
$a$ around 1 GeV and using an average energy $E_\gamma^{\rm LAB}=3.8$
GeV.  First, the figure shows that our election for the Regge slope
$\alpha_{K}'(0)\approx 0.8$ [GeV$^{-2}$] provides a fair description
of the $t-$dependence of the differential cross section when the
gaussian cutoff is fixed to values around 1 GeV. Second, we see a mild
dependence of the results on $a$, and find that the use of $a= 1$ GeV
provides a slightly better description of the $t-$dependence of the
differential cross section.

Hybrid hadron/$K-$Reggeon exchange model results for the $\gamma p
\rightarrow K^+ \Lambda^*$ total cross section are shown in
Fig.~\ref{fig:figtl} from threshold up to photon energies (in the LAB
frame) around 5 GeV.  We always neglect the $K^*-$contributions, both
in the hadron and in the quark-gluon string model. In the effective
hadron Lagrangian model, we neglect the terms affected by the $f_s$
form-factor, since they are greatly suppressed by it~\cite{nam05}. We
have examined two different values of the cutoff $\Lambda$ entering in
the hadron form-factors of Eq.~(\ref{eq:cut-had}), and as can be
appreciated in the figure, the predictions depend drastically on the
precise used value. For the weighting function of
Eq~(\ref{eq:weight}), which characterizes the hybrid model, the values
$E_0~=~2.3$\,GeV (for $\Lambda=0.75$ GeV) and 2.0\,GeV (for
$\Lambda=1$ GeV) and $\Delta E~=~0.1$\,GeV have been used.

We also show, in Fig.~\ref{fig:figtl}, results from the the $K$ Reggeon
exchange model in the entire photon energy range. This latter reaction
mechanism leads to cross sections of about $2 \mu$b at
$E_\gamma^{\rm LAB}=$2 GeV.  It would be highly desirable to count
with experimental measurements of the cross section around these
energies, for which occur the transition between the hadron and the string
quark-gluon reaction mechanisms~\cite{nakano07}.  

Finally, in Fig.~\ref{fig:figad} we present results for the $E^{\rm
  LAB}_\gamma=$2 GeV cm angular distributions from the three models
studied in Fig.~\ref{fig:figtl}. For all cases the differential cross
section peaks forward and gradually fall down as the cm angle
increases.

\section{Conclusion}
\label{sec:concl}

We have studied the photon induced $\Lambda^*$ production using both
an effective hadron model and a quark-gluon string approach, in a wide range
of laboratory photon energies from threshold up to 5 GeV.  

We have studied first the coupling constant $g_{K^*N\Lambda^*}$ within
the SU(6) chiral unitary model proposed in Ref.~\cite{garcia06-su6},
and found that this coupling constant is of the order of 0.5. This
value is much smaller than both the value used by phenomenological
studies and that of $g_{K N\Lambda^*}$, determined by the $\Lambda^*$
decay width.  Hence, we conclude that the contribution of $t-$channel
$K^*$ exchange in the effective Lagrangian approach can be safely
neglected. We have also discussed the existing connection between the
small value found for $g_{K^*N\Lambda^*}$ and the proximity of the
$\Lambda^*$ mass to the $\pi \Sigma^*$ threshold.

We have also shown that the quark-gluon string reaction mechanism,
realized in the Reggeon exchange model, is able to reproduce the
available experimental data in the region from $E^{\rm LAB}_\gamma
\sim 2.8$ GeV up to 5 GeV.  Here again, we find that the
$K-$trajectory, with a 1 GeV cutoff parameter for the gaussian
form-factor, reasonably describes the energy and angular dependence of
the cross section. 

We should also mention that Titov advocated for the very first time a
quark-gluon string approach to study the $\gamma p \to K^+ \Lambda^* $
reaction~\cite{titov05}. However, he assumed a clear dominance of the
$K^*$ $t-$exchange contribution in the hadron approach near threshold
and to describe the higher energy region, where the data lie, he used
a reggezation of the $K^*$ meson propagator.  We would like to make
here two remarks. First, our results for the $K^*N \Lambda^*$ coupling
contradict Titov's assumption that the $K^*$ $t-$exchange 
is the dominant mechanism near threshold (he
used a value for this coupling around a factor 10 larger than that
deduced here). Second, and as we have already mentioned, the
$K^*-$trajectory used in this reference does not lead to
$\alpha_{K^*}(m_{K^*}^2)=1$.  Indeed, it is much closer to our
$K-$trajectory than to the $K^*-$one in the $t-$range where the data
have been measured.

We have smoothly extended this Reggeon exchange model down to smaller
energies and found that it leads to cross sections of around 2 $\mu$b
in the $E^{\rm LAB}_\gamma \sim 2$ GeV region.  We have then proposed
a hybrid model to connect with the meson-baryon approach. Finally, we
have shown that the effective Lagrangian model predictions depend
largely on the hadron form-factor.  It would be very important to
compare with experimental measurements of the cross sections in the
energy region of $E^{\rm LAB}_\gamma \sim 2$ GeV in order to determine
this form-factor and also to better understand the transition between
the meson-baryon and the quark-gluon mechanisms.

\vspace{2cm}

H.T.  acknowledges the hospitality of the Departamento de F\'isica
At\'omica, Molecular y Nuclear, Universidad de Granada, MEC support
for {\it Estancias de Profesores e investigadores extranjeros en r\'egimen
de a\~no sab\'atico en Espa\~na} SAB2006--0150 and the support by
the Grant-in-aid for Scientific Research (C)18540269 of
the Ministry of Education in Japan.
CGR and JN acknowledges support from Junta de Andalucia grant
FQM0225, MEC grant FIS2005--00810 and
from the EU Human Resources and Mobility Activity, FLAVIAnet, contract
number MRTN--CT--2006--035482.

\end{document}